\documentclass[twocolumn,aps,prb,showpacs]{revtex4}
\usepackage{graphicx,epsfig,amssymb}
\usepackage{amsmath}
\usepackage{graphicx}
\usepackage{amsfonts}
\usepackage{amssymb}
\usepackage{color}
\usepackage{ulem}
\begin{document}

\textbf{Comment on "Apical charge flux-modulated in-plane transport properties
of cuprate superconductors"}\bigskip

The abstract of Ref.$\,[$1] claims to "demonstrate, using \textit{ab initio}
computations, a new trend suggesting that the cuprates with stronger
out-of-CuO$_{2}$-plane chemical bonding between the apical anion (O, Cl) and
apical cation (e.g., La, Hg, Bi, Tl) are generally correlated with higher
$T_{c\,\max }$ in experiments". We point out that this trend is \textit{included} in the long-known [2] correlation of $T_{c\,\max }$ with the hopping range 
$\left( \sim t^{\prime }/t\right) $ of the electrons at (the most interlayer-bonding sheet of) the Fermi-surface.
Contrary to the impression given in Ref.$\,[1]$, the correlation [2]
is not simply with the distance, $d_{A},$ of apical oxygen from the nearest
CuO$_{2}$ plane; but rather, as stated in the abstract of Ref.$\,$[2], "It is
controlled by the energy of the axial orbital, a hybrid between Cu 4$s$,
apical-oxygen 2$p_{z},$ and farther orbitals." The bonding between the apical
cation (AC) and apical anion (AA) describing  $T_{c\text{ }\max }$ in
Ref.$[1]$ is included in the energy of the axial orbital.

The axial orbital ($s$), is the 4th orbital which must
be added to the nearest-neighbor (n.n.) $\,$3-band CuO$_{2}$ model,
parametrized by $\varepsilon _{d},$ $\varepsilon _{p},$ and $t_{pd}$,
to define a 4-band model with two additional
parameters: the energy, $\varepsilon _{s},$ of the axial orbital and its
hopping integral, $t_{sp},$ to a n.n.$\,$oxygen in the CuO$_{2}$ plane
(FIG.1 in Ref.[2]). The axial orbital is the vehicle for hopping around a
corner of the square Cu lattice, a hopping which in the 2.n.n.$\,$3-band
model is described $t_{pp}=t_{sp}^{2}\left/  \left(  \varepsilon_{s}-\varepsilon_{F}\right)
\right.  ,$ and in the one-band model by: $t^{\prime}\approx rt.$
Here $r=$ $2t_{pp}/\left( \varepsilon_{F}-\varepsilon _{p}+4t_{pp}\right)$
is the range parameter. Of the five parameters of the n.n.$\,$4-band model, 
\textit{only} $\varepsilon_{s}$ was found to vary significantly among
materials. Hence, the most microscopic parameter found to correlate
positively with $T_{c\,\max }$ was the inverse of the excitation energy, $\varepsilon _{s}-\varepsilon _{F},$ for hopping around a corner.

The axial orbital (FIG.$\,$3 and Eq.(3) plus the two following lines in Ref.$\,$[2]) is the Cu 4$s$-\textit{like} hybrid
between the axial atomic orbitals on Cu, the two apical
oxygens (AA), and farther orbitals, such as La 5$d_{3z^{2}-1}$ and Tl
6$p_{z},$ on the two apical cations (AC). In terms of the energies, $\varepsilon _{\mathrm{AA}}<\varepsilon _{F}<\varepsilon _{\mathrm{Cu}\,4s}<\varepsilon _{\mathrm{AC}}$, of, and hopping integrals, $t_{%
\mathrm{AA,\,Cu}\,4s}$ and $t_{\mathrm{AA,\,\,AC}},$ between these \textit{atomic} 
constituents, the energy of the axial orbital is:%
\begin{equation}
\varepsilon _{s}=\varepsilon _{\mathrm{Cu}\,4s}+\frac{2t_{\mathrm{AA,\,Cu}%
\,4s}^{2}}{\varepsilon _{\mathrm{Cu}\,4s}-\left( \varepsilon _{\mathrm{AA}}-%
\frac{t_{\mathrm{AA,\,\,AC}}^{2}}{\varepsilon _{\mathrm{AC}}-\varepsilon _{%
\mathrm{AA}}}\right) }.  \label{es}
\end{equation}%
Here, the energy, $\varepsilon _{\mathrm{AA}},$ of apical-oxygen$\,$2$p_{z}$
is lowered due to the repulsion from the AC orbital and the energy of Cu$\,$4%
$s$ is raised due to the repulsion from apical-oxygen$\,$2$p_{z}$. The
bonding between AA and AC, demonstrated in Ref.$\,$[1] to correlate
positively with $T_{c\,\max },$ is essentially the amount, $t_{\mathrm{%
AA,\,AC}}^{2}/\left( \varepsilon _{\mathrm{AC}}-\varepsilon _{\mathrm{AA}%
}\right) ,$ by which the denominator in Eq.$\,$(\ref{es}) is increased due
to this bond. This, in turn, is seen to cause the energy of the axial
orbital to decrease towards that of the bare Cu 4$s$ and, hence, according
to Ref.$\,$[2] be associated with a higher $T_{c\,\max };$ Q.E.D.! In the words
of Ref. [1]: "..$\,$the closer the hybridization peak is to the Fermi level $%
\left( \varepsilon _{F}\right) $, the higher the $T_{c\,\max }$ is."

Eq.$\,($\ref{es}) also exhibits other causes for an increased $T_{c\,\max }$
captured by [2], but not by [1]. For instance, increasing the
distance, $d_{A},$ to AA, thus \textit{de}creasing the AA-AC distance, causes not only
an increase of $t_{\mathrm{AA,\,AC}}^{2},$ but also a decrease of
$t_{\mathrm{AA,\,Cu\,4s}}^{2}$, and thereby leads to an even stronger decrease of $\varepsilon _{s}$
towards $\varepsilon _{\mathrm{Cu}\,4s}$ and, hence, stronger increase of $T_{c\,\max }.$ That $T_{c\,\max }$ generally increases with $d_{A}$ is, however, \textit{not} true, because also $\varepsilon _{AA}$ and $\varepsilon _{\mathrm{%
AC}}$ (the nature of the AA and the AC) matter, unless $d_{A}$ is so large
that $t_{\mathrm{AA,\,Cu\,4s}}^{2}\mathrm{\sim }0$ (almost the case of Tl
and Hg in FIG.$\,$4 of Ref.$\,$[2]).

Most importantly, whereas the
correlation [2] extends in a straight-forward way to the numerous higher-$%
T_{c}$ bi- and triplelayer cuprates (FIG.$\,$5 in Ref.$\,$[2]), the
correlation [1] does \textit{not}, simply because the bonds between AA and AC
are \textit{outside} the multilayer.  Instead, Ref.$\,$%
[1] ascribes the observed trend to charge transfer from a vertically
vibrating AA to a CuO$_{2}$ layer, i.e. a higher-order, dynamical process.
This, we feel, deserves further study.

\bigskip 

E. Pavarini$^{1}$, I. Dasgupta$^{2}$, T. Saha-Dasgupta$^{2}$, and O.K.
Andersen$^{3,\ast }$

$^{1}$Forschungszentrum Julich, Inst Adv Simulat, D-52425 Julich, Germany

$^{2}$School of Physical Sciences, Indian Assoc Cultivat Sci, Kolkata 700
032, India

$^{3\ast }$Max Planck Inst Solid State Res, Heisenbergstr 1, D-70569
Stuttgart, Germany. oka@fkf.mpg.de

\bigskip 

[1] S. Kim, X. Chen, W. Fitzhugh, and X. Li, Phys. Rev. Lett. \textbf{121},%
\textbf{\ }157001 (2018).

[2] E. Pavarini, I. Dasgupta, T. Saha-Dasgupta, O. Jepsen, and O.K.
Andersen, Phys. Rev. Lett. \textbf{87}, 047003 (2001).

\end{document}